%% file: ms.tex
\documentclass[a4paper]{article}

\usepackage[english]{babel}
\usepackage[utf8]{inputenc}
\usepackage[T1]{fontenc}

\usepackage[a4paper,top=3cm,bottom=2cm,left=3cm,right=3cm,marginparwidth=1.75cm]{geometry}

\usepackage[tbtags]{amsmath}
\usepackage{mathtools}
\usepackage{graphicx}
\usepackage[colorinlistoftodos]{todonotes}
\usepackage[colorlinks=true, allcolors=blue]{hyperref}
\usepackage{color}
\usepackage{tikz}
\usepackage{titlesec}
\usepackage{caption}
\usepackage[shortlabels]{enumitem}
\usepackage{setspace}
\usepackage{authblk}
\usepackage{floatrow}

\floatstyle{plaintop}
\restylefloat{table}

\usetikzlibrary{shapes.geometric, arrows}

\usepackage[backend=bibtex,style=authoryear-comp,sorting=nyt,isbn=false,url=false, natbib=true, doi=false]{biblatex}
\tikzstyle{startstop_big} = [rectangle, rounded corners, minimum width=3cm, minimum height=1cm,text centered, draw=black,text width = 8.5cm, fill=red!30]
\tikzstyle{startstop} = [rectangle, rounded corners, minimum width=3cm, minimum height=1cm,text centered, draw=black,text width = 5cm, fill=red!30]
\tikzstyle{io} = [trapezium, trapezium left angle=70, trapezium right angle=110, minimum width=2cm, minimum height=1cm, text centered, draw=black, fill=blue!30]
\tikzstyle{process} = [rectangle, minimum width=3cm, minimum height=1cm, text centered, text width = 3cm, draw=black, fill=orange!30]
\tikzstyle{process_wide_3pt5cm} = [rectangle, minimum width=3cm, minimum height=1cm, text centered, text width = 3.5cm, draw=black, fill=orange!30]
\tikzstyle{process_wide_4cm} = [rectangle, minimum width=3cm, minimum height=1cm, text centered, text width = 4cm, draw=black, fill=orange!30]
\tikzstyle{process_wide_4pt5cm} = [rectangle, minimum width=3cm, minimum height=1cm, text width = 4.5cm, draw=black, fill=orange!30]
\tikzstyle{process_wide_5cm} = [rectangle, minimum width=3cm, minimum height=1cm, text width = 5cm, draw=black, fill=orange!30]
\tikzstyle{decision} = [diamond, minimum width=1cm, minimum height=1cm,text centered,text width = 2.25cm, draw=black, fill=green!30]
\tikzstyle{arrow} = [thick,->,>=stealth]

\definecolor{ErasmusBlue}{RGB}{12, 32, 116}

\DeclareMathOperator{\argmax}{arg\,max}

\let\bmath\boldsymbol
\let\rmn\mathrm
\newcommand{\Hline}
{
\hline
\hline
}

\begin{document}
\title{\textbf{Personalized Schedules for Surveillance of Low Risk Prostate Cancer Patients}}

\author[1,*]{Anirudh Tomer}
\author[2]{Daan Nieboer }
\author[3]{Monique J. Roobol }
\author[2,4]{Ewout W. Steyerberg }
\author[1]{Dimitris Rizopoulos}
\affil[1]{Department of Biostatistics, Erasmus University Medical Center, the Netherlands}
\affil[2]{Department of Public Health, Erasmus University Medical Center, the Netherlands}
\affil[3]{Department of Urology, Erasmus University Medical Center, the Netherlands}
\affil[4]{Department of Medical Statistics and Bioinformatics, Leiden University Medical Center, the Netherlands}
\affil[ ]{*\textit {email}: a.tomer@erasmusmc.nl}

\date{}

\maketitle

\input{mainmatter/abstract}
%\doublespacing
\input{mainmatter/introduction}
\input{mainmatter/joint_model_framework}
\input{mainmatter/pers_sched_framework/pers_biopsy_framework}
\input{mainmatter/choosing_schedule}
\input{mainmatter/pers_schedule_prias}
\input{mainmatter/sim_study}
\input{mainmatter/discussion}
\input{mainmatter/acknowledgement}

\section*{Supplementary Materials}

Web Appendix A, C, D and E referenced in Section~\ref{sec : jm_framework},  Section~\ref{sec : pers_schedule_PRIAS},  and  Section~\ref{sec: discussion}, respectively, and the derivation of Equation (\ref{eq : expected_time_survprob}) and (\ref{eq : var_time_survprob}) in Web Appendix B, are available in a supplementary document uploaded at \url{https://goo.gl/jpPtL8}.\vspace*{-8pt}

\textbf{}

\clearpage
\printbibliography

\appendix

\end{document}

%% file: mainmatter/abstract.tex
% !TEX root =  ../pers_schedules.tex 

\begin{abstract}
Low risk prostate cancer patients enrolled in active surveillance (AS) programs commonly undergo biopsies on a frequent basis for examination of cancer progression. AS programs employ a fixed schedule of biopsies for all patients. Such fixed and frequent schedules, may schedule unnecessary biopsies for the patients. Since biopsies have an associated risk of complications, patients do not always comply with the schedule, which increases the risk of delayed detection of cancer progression. Motivated by the world's largest AS program, Prostate Cancer Research International Active Surveillance (PRIAS), in this paper we present personalized schedules for biopsies to counter these problems. Using joint models for time to event and longitudinal data, our methods combine information from historical prostate-specific antigen (PSA) levels and repeat biopsy results of a patient, to schedule the next biopsy. We also present methods to compare personalized schedules with existing biopsy schedules.
\end{abstract}

%% file: mainmatter/introduction.tex
% !TEX root =  ../pers_schedules.tex 
\section{Introduction}
\label{sec : introduction}
Cancer screening is a widely used practice to detect cancer before symptoms appear in otherwise healthy individuals. A major issue of screening programs that has been observed in many types of cancers is overdiagnosis \citep{esserman2014addressing}. To avoid subsequent overtreatment, patients diagnosed with low grade cancers are commonly advised to join active surveillance (AS) programs. The goal of AS is to routinely examine the progression of cancer and avoid serious treatments such as surgery, chemotherapy, or radiotherapy as long as they are not needed. To this end, AS includes, but is not limited to periodical evaluation of biomarkers pertaining to the cancer, physical examination, medical imaging, and biopsy.

In this paper we focus on AS programs for prostate cancer (PCa), wherein the decision to exit AS and start active treatment (e.g., operation, chemotherapy) is typically based on invasive examinations, such as biopsies \citep{bokhorst2016decade}. Biopsies can be reliable, but they are also painful, and have an associated risk of complications such as urinary retention, hematuria and sepsis \citep{loeb2013systematic}. Because of this reason the schedule of  biopsies has significant medical consequences for patients. A frequent schedule of biopsies may help detecting PCa progression earlier but the corresponding risk of complications will be high. Although such a schedule may work well for patients with faster progressing cancer, for slowly progressing PCa patients many unnecessary biopsies may be scheduled. Furthermore, patients do not always comply with such a schedule, as has been observed by \citet{bokhorst2015compliance}. In the specific case of the world's largest AS program, Prostate Cancer Research International Active Surveillance (PRIAS) \citep{bokhorst2016decade}, the compliance rate for biopsies steadily decreased from 81\% at year one of follow up, to 60\% at year four, 53\% at year seven and 33\% at year ten. Such non-compliance can lead to delayed detection of PCa progression, which may reduce the effectiveness of AS programs.

This paper is motivated by the need to reduce the medical burden of repeat biopsies while simultaneously avoiding late detection of PCa progression. For the latter purpose, several AS programs employ a fixed annual schedule (biopsies with a gap of one year) of biopsies \citep{tosoian2011active,welty2015extended}. However, given the medical burden of biopsies, most AS programs also strongly advise against scheduling biopsies more frequently than the annual schedule. The PRIAS schedule for biopsies is relatively lenient: one biopsy each is scheduled at year one of follow up, year four, year seven, year ten, and every five years thereafter. However, PRIAS also switches to the annual schedule if a patient's prostate-specific antigen (PSA) doubling time, also known as PSA-DT and measured as the inverse of the slope of the regression line through the base two logarithm of PSA values, is less than 10 years. We intend to improve upon such fixed schedules by creating personalized schedules for biopsies. That is, a different schedule for every patient utilizing their periodically measured serum PSA levels (measured in ng/mL) and repeat biopsy results. Biopsies are graded using the Gleason score, which takes an integer value between 6 and 10, with 10 corresponding to the most serious state of the cancer. Patients enter AS only if their Gleason score is 6. When the Gleason score becomes greater than 6, also known as Gleason reclassification (referred to as GR hereafter), patients are often advised to switch from AS to active treatment. Hence, for AS programs it is of prime interest to detect GR soonest with least number of biopsies possible.

Personalized schedules for screening have received much interest in the literature, especially in the medical decision making context. For diabetic retinopathy, cost optimized personalized schedules based on Markov models have been developed by \citet{bebu2017OptimalScreening}. For breast cancer, personalized mammography screening policy based on the prior screening history and personal risk characteristics of women, using partially observable Markov decision process (MDP) models have been proposed by \citet*{ayer2012or}. MDP models have also been used to develop personalized screening policies for cervical cancer \citep*{akhavan2017markov} and colorectal cancer \citep*{erenay2014optimizing}. Another type of model called joint model for time to event and longitudinal data \citep{tsiatis2004joint,rizopoulos2012joint} has also been used to create personalized schedules, albeit for the measurement of longitudinal biomarkers \citep{drizopoulosPersScreening}. In context of PCa, \citet{zhang2012optimization} have used partially observable MDP models to personalize the decision of (not) deferring a biopsy to the next checkup time during the screening process. The decision is based on the baseline characteristics as well as a discretized PSA level of the patient at the current check up time.

Our work differs from the above referenced work in certain aspects. Firstly, the schedules we propose in this paper, account for the latent between-patient heterogeneity. We achieve this using joint models, which are inherently patient-specific because they utilize random effects. Secondly, joint models allow a continuous time scale and utilize the entire history of PSA levels. Lastly, instead of making a binary decision of (not) deferring a biopsy to the next pre-scheduled check up time, we schedule biopsies at a per patient optimal future time. To this end, using joint models we first obtain a full specification of the joint distribution of PSA levels and time of GR. We then use it to define a patient-specific posterior predictive distribution of the time of GR given the observed PSA measurements and repeat biopsies up to the current check up time. Using the general framework of Bayesian decision theory, we propose a set of loss functions which are minimized to find the optimal time of conducting a biopsy. These loss functions yield us two categories of personalized schedules, those based on expected time of GR and those based on the risk of GR. In addition we analyze an approach where the two types of schedules are combined. We also present methods to evaluate and compare the various schedules for biopsies.

The rest of the paper is organized as follows. Section \ref{sec : jm_framework} briefly covers the joint modeling framework. Section \ref{sec : pers_sched_approaches} details the personalized scheduling approaches we have proposed in this paper. In Section \ref{sec : choosing_schedule} we discuss methods for evaluation and selection of a schedule. In Section \ref{sec : pers_schedule_PRIAS} we demonstrate the personalized schedules by employing them for the patients from the PRIAS program. Lastly, in Section \ref{sec: simulation_study}, we present the results from a simulation study we conducted to compare personalized schedules with PRIAS and annual schedule.

%% file: mainmatter/joint_model_framework.tex
% !TEX root =  ../pers_schedules.tex 
\section{Joint Model for Time to Event and Longitudinal Outcomes}
\label{sec : jm_framework}
We start with the definition of the joint modeling framework that will be used to fit a model to the available dataset, and then to plan biopsies for future patients. Let $T_i^*$ denote the true GR time for the $i$-th patient enrolled in an AS program. Let $S$ be the schedule of biopsies prescribed to this patient. The corresponding vector of time of biopsies is denoted by $T_i^S = \{T^S_{i0}, T^S_{i1}, \ldots, T^S_{i{N_i^S}}; T^S_{ij} < T^S_{ik}, \forall j<k\}$, where $N_i^S$ are the total number of biopsies conducted. Because of the periodical nature of biopsy schedules, $T_i^*$ cannot be observed directly and it is only known to fall in an interval $l_i < T_i^* \leq r_i$, where $l_i = T^S_{i{N_i^S - 1}}, r_i = T^S_{i{N_i^S}}$ if GR is observed, and $l_i = T^S_{i{N_i^S}}, r_i=\infty$ if GR is not observed yet. Further let $\bmath{y}_i$ denote the $n_i \times 1$  vector of PSA levels for the $i$-th patient. For a sample of $n$ patients the observed data is denoted by $\mathcal{D}_n = \{l_i, r_i, \bmath{y}_i; i = 1, \ldots, n\}$.

The longitudinal outcome of interest, namely PSA level, is continuous in nature and thus to model it the joint model utilizes a linear mixed effects model (LMM) of the form:
\begin{equation*}
\begin{split}
y_i(t) &= m_i(t) + \varepsilon_i(t)\\
&=\bmath{x}_i^T(t) \bmath{\beta} + \bmath{z}_i^T(t) \bmath{b}_i + \varepsilon_i(t),
\end{split}
\end{equation*}
where $\bmath{x}_i(t)$ denotes the row vector of the design matrix for fixed effects and $\bmath{z}_i(t)$ denotes the same for random effects. Correspondingly the fixed effects are denoted by $\bmath{\beta}$ and random effects by $\bmath{b}_i$. The random effects are assumed to be normally distributed with mean zero and $q \times q$ covariance matrix $\bmath{D}$. The true and unobserved PSA level at time $t$ is denoted by $m_i(t)$. Unlike $y_i(t)$, the former is not contaminated with the measurement error $\varepsilon_i(t)$. The error is assumed to be normally distributed with mean zero and variance $\sigma^2$, and is independent of the random effects $\bmath{b}_i$.

To model the effect of PSA on hazard of GR, joint models utilize a relative risk sub-model. The hazard of GR for patient $i$ at any time point $t$, denoted by $h_i(t)$, depends on a function of subject specific linear predictor $m_i(t)$ and/or the random effects:
\begin{align*}
h_i(t \mid \mathcal{M}_i(t), \bmath{w}_i) &= \lim_{\Delta t \to 0} \frac{\mbox{Pr}\big\{t \leq T^*_i < t + \Delta t \mid T^*_i \geq t, \mathcal{M}_i(t), \bmath{w}_i\big\}}{\Delta t}\\
&=h_0(t) \exp\big[\bmath{\gamma}^T\bmath{w}_i + f\{M_i(t), \bmath{b}_i, \bmath{\alpha}\}\big], \quad t>0,
\end{align*}
where $\mathcal{M}_i(t) = \{m_i(v), 0\leq v \leq t\}$ denotes the history of the underlying PSA levels up to time $t$. The vector of baseline covariates is denoted by $\bmath{w}_i$, and $\bmath{\gamma}$ are the corresponding parameters. The function $f(\cdot)$ parametrized by vector $\bmath{\alpha}$ specifies the functional form of PSA levels \citep{brown2009assessing,rizopoulos2012joint,taylor2013real,rizopoulos2014bma} that is used in the linear predictor of the relative risk model. Some functional forms relevant to the problem at hand are the following: 
\begin{eqnarray*}
\left \{
\begin{array}{l}
f\{M_i(t), \bmath{b}_i, \bmath{\alpha}\} = \alpha m_i(t),\\
f\{M_i(t), \bmath{b}_i, \bmath{\alpha}\} = \alpha_1 m_i(t) + \alpha_2 m'_i(t),\quad \text{with}\  m'_i(t) = \frac{\rmn{d}{m_i(t)}}{\rmn{d}{t}}.\\
\end{array}
\right.
\end{eqnarray*}
These formulations of $f(\cdot)$ postulate that the hazard of GR at time $t$ may be associated with the underlying level $m_i(t)$ of the PSA at $t$, or with both the level and velocity $m'_i(t)$ of the PSA at $t$. Lastly, $h_0(t)$ is the baseline hazard at time $t$, and is modeled flexibly using P-splines. More specifically:
\begin{equation*}
\log{h_0(t)} = \gamma_{h_0,0} + \sum_{q=1}^Q \gamma_{h_0,q} B_q(t, \bmath{v}),
\end{equation*}
where $B_q(t, \bmath{v})$ denotes the $q$-th basis function of a B-spline with knots $\bmath{v} = v_1, \ldots, v_Q$ and vector of spline coefficients $\gamma_{h_0}$. To avoid choosing the number and position of knots in the spline, a relatively high number of knots (e.g., 15 to 20) are chosen and the corresponding B-spline regression coefficients $\gamma_{h_0}$ are penalized using a differences penalty \citep{eilers1996flexible}. 

For the estimation of joint model's parameters we use a Bayesian approach. The details of the estimation method are presented in Web Appendix A of the supplementary material.

%% file: mainmatter/pers_sched_framework/pers_biopsy_framework.tex
% !TEX root =  ../pers_schedules.tex 
\section{Personalized Schedules for Repeat Biopsies}
\label{sec : pers_sched_approaches}
Once a joint model for GR and PSA levels is obtained, the next step is to use it to create personalized schedules for biopsies. Let us assume that a personalized schedule is to be created for a new patient $j$, who is not present in the original sample $\mathcal{D}_n$ of patients. Further let us assume that this patient did not have a GR at his last biopsy performed at time $t$, and that the PSA levels are available up to a time point $s$. The goal is to find the optimal time $u > \mbox{max}(t,s)$ of the next biopsy. 

\input{mainmatter/pers_sched_framework/ppd_time_to_GR}
\input{mainmatter/pers_sched_framework/loss_functions}
\input{mainmatter/pers_sched_framework/estimation}

\subsection{Algorithm}
\label{subsec : pers_sched_algorithm}
\begin{figure}
\centerline{\input{mainmatter/pers_sched_framework/algorithm}}
\caption{Algorithm for creating a personalized schedule for patient $j$. The time of the latest biopsy is denoted by $t$. The time of the latest available PSA measurement is denoted by $s$. The proposed personalized time of biopsy is denoted by $u$.  The time at which a repeat biopsy was proposed on the last visit to the hospital is denoted by $u^{pv}$. The time of the next visit for the measurement of PSA is denoted by $T^{nv}$.} 
\label{fig : sched_algorithm}
\end{figure}
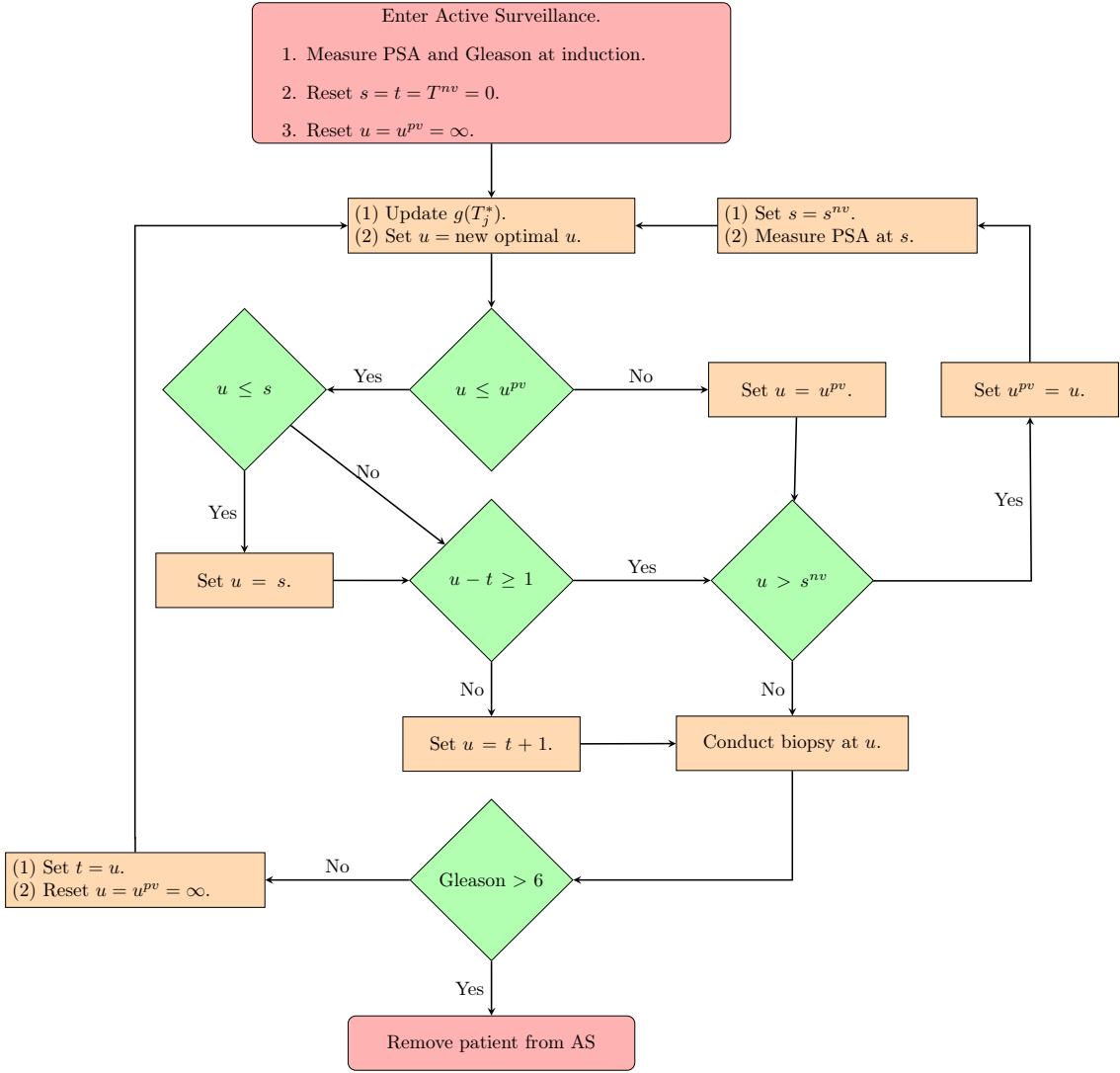

The aforementioned personalized schedules, schedule biopsy at a time $u > \mbox{max}(t,s)$. However, if time $u < T^*_j$, then GR is not detected at $u$ and at least one more biopsy is required at an optimal time $u^{new} > \mbox{max}(u,s)$. This process is repeated until GR is detected. To aid in medical decision making, we elucidate this process via an algorithm in Figure \ref{fig : sched_algorithm}. Since AS programs strongly advise that biopsies are conducted at a gap of at least one year, when $u - t < 1$, the algorithm postpones $u$ to $t + 1$, because it is the time nearest to $u$, at which the one year gap condition is satisfied.

%When submitting replace figure with figure* to make it span an entire column
%With referee option it gives an error

%\begin{figure*}
%\centerline{\input{mainmatter/pers_sched_framework/algorithm_two_col}}
%\caption{Algorithm for creating a personalized schedule for patient $j$. $t$ denotes the time of the latest biopsy. $s$ denotes the time of the latest available PSA measurement. $u$ denotes the proposed personalized time of biopsy.  $u^{pv}$ denotes the time at which a repeat biopsy was proposed on the last visit to the hospital. $T^{nv}$ denotes the time of the next visit for measurement of PSA.} 
%\label{fig : sched_algorithm}
%\end{figure*}

%% file: mainmatter/pers_sched_framework/ppd_time_to_GR.tex
% !TEX root =  ../pers_schedules.tex 

\subsection{Posterior Predictive Distribution for Time to GR}
\label{subsec : ppd_time_to_GR}
Let $\mathcal{Y}_j(s)$ denote the history of PSA levels taken up to time $s$ for patient $j$. The information from PSA history and repeat biopsies is manifested by the posterior predictive distribution $g(T^*_j)$, given by (conditioning on baseline covariates $\bmath{w}_i$ is dropped for notational simplicity hereafter):
\begin{equation}
\label{eq : dyn_dist_fail_time}
\begin{split}
g(T^*_j) &= p\big\{T^*_j \mid T^*_j > t, \mathcal{Y}_j(s), \mathcal{D}_n\big\}\\
&= \int p\big\{T^*_j \mid T^*_j > t, \mathcal{Y}_j(s), \bmath{\theta}\big\}p\big(\bmath{\theta} \mid \mathcal{D}_n\big) \rmn{d} \bmath{\theta}\\
&= \int \int p\big\{T^*_j \mid T^*_j > t, \bmath{b_j}, \bmath{\theta}\big\}p\big\{\bmath{b}_j \mid T^*_j>t, \mathcal{Y}_j(s), \bmath{\theta}\big\}p\big(\bmath{\theta} \mid \mathcal{D}_n\big) \rmn{d} \bmath{b}_j \rmn{d} \bmath{\theta}.
\end{split}
\end{equation}
The distribution $g(T^*_j)$ depends on the observed longitudinal history $\mathcal{Y}_j(s)$ of patient $j$ via the random effects $\bmath{b_j}$, and on the information from the original dataset $\mathcal{D}_n$ via the posterior distribution of the parameters $p(\bmath{\theta} \mid \mathcal{D}_n)$, where $\bmath{\theta}$ denotes the vector of all parameters.

%% file: mainmatter/pers_sched_framework/loss_functions.tex
% !TEX root =  ../pers_schedules.tex 

\subsection{Loss Functions}
\label{subsec : loss_functions}
To find the time $u$ of the next biopsy, we use principles from statistical decision theory in a Bayesian setting \citep{bergerDecisionTheory,robertBayesianChoice}. More specifically, we propose to choose $u$ by minimizing the posterior expected loss $E_g\big\{L(T^*_j, u)\big\}$, where the expectation is taken with respect to $g(T^*_j)$. The former is given by:
\begin{equation*}
E_g\big\{L(T^*_j, u)\big\} = \int_t^\infty L(T^*_j, u) p\big\{T^*_j \mid T^*_j > t, \mathcal{Y}_j(s), \mathcal{D}_n\big\} \rmn{d} T^*_j.
\end{equation*}
Various loss functions $L(T^*_j, u)$ have been proposed in literature \citep{robertBayesianChoice}. The ones we utilize, and the corresponding motivations are presented next.

Given the medical burden of biopsies, ideally only one biopsy performed at the exact time of GR is sufficient. Hence, neither a time which overshoots the true GR time $T^*_j$, nor a time which undershoots is preferred. In this regard, the squared loss function $L(T^*_j, u) = (T^*_j - u)^2$ and the absolute loss function $L(T^*_j, u) = \left|{T^*_j - u}\right|$ have the properties that the posterior expected loss is symmetric on both sides of $T^*_j$. Secondly, both loss functions have well known solutions available. The posterior expected loss for the squared loss function is given by:
\begin{equation}
\label{eq : posterior_squared_loss}
\begin{split}
E_g\big\{L(T^*_j, u)\big\} &= E_g\big\{(T^*_j - u)^2\big\}\\
&=E_g\big\{(T^*_j)^2\big\} + u^2 -2uE_g(T^*_j).
\end{split}
\end{equation}
The posterior expected loss in (\ref{eq : posterior_squared_loss}) attains its minimum at $u = E_g(T^*_j)$, the expected time of GR. The posterior expected loss for the absolute loss function is given by:
\begin{equation}
\label{eq : posterior_absolute_loss}
\begin{split}
E_g\big\{L(T^*_j, u)\big\} &= E_g\big(\left|{T^*_j - u}\right|\big)\\
&= \int_u^\infty (T^*_j - u) g(T^*_j)\rmn{d} T^*_j + \int_t^u (u - T^*_j) g(T^*_j)\rmn{d} T^*_j.
\end{split}
\end{equation}
The posterior expected loss in (\ref{eq : posterior_absolute_loss}) attains its minimum at the median of $g(T^*_j)$, given by $u = \pi_j^{-1}(0.5 \mid t,s)$, where $\pi_j^{-1}(\cdot)$ is the inverse of dynamic survival probability $\pi_j(u \mid t, s)$ of patient $j$ \citep{rizopoulos2011dynamic}. It is given by:
\begin{equation}
\label{eq : dynamic_surv_prob}
\pi_j(u \mid t, s) = \mbox{Pr}\big\{T^*_j \geq u \mid  T^*_j >t, \mathcal{Y}_j(s), D_n\big\}, \quad u \geq t.
\end{equation}
For ease of readability we denote $\pi_j^{-1}(0.5 \mid t,s)$ as $\mbox{median}(T^*_j)$ hereafter.

Even though the mean or median time of GR may be obvious choices from a statistical perspective, from the viewpoint of doctors or patients, it could be more intuitive to make the decision for the next biopsy by placing a cutoff $1 - \kappa$, where $0 \leq \kappa \leq 1$, on the dynamic incidence/risk of GR. This approach would be successful if $\kappa$ can sufficiently well differentiate between patients who will obtain GR in a given period of time, and those who will not. This approach is also useful when patients are apprehensive about delaying biopsies beyond a certain risk cutoff. Thus, a biopsy can be scheduled at a time point $u$ such that the dynamic risk of GR is higher than a certain threshold $1 - \kappa,\ $ beyond $u$. To this end, the posterior expected loss for the following multilinear loss function can be minimized to find the optimal $u$:
\begin{equation}
\label{eq : loss_dynamic_risk}
L_{k_1, k_2}(T^*_j, u) =
    \begin{cases}
      k_2(T^*_j-u), k_2>0 & \text{if } T^*_j > u,\\
      k_1(u-T^*_j), k_1>0 & \text{otherwise}.
    \end{cases}       
\end{equation}
where $k_1, k_2$ are constants parameterizing the loss function. The posterior expected loss $E_g\big\{L_{k_1, k_2}(T^*_j, u)\big\}$ obtains its minimum at $u = \pi_j^{-1}\big\{k_1/{(k_1 + k_2)} \mid t,s \big\}$ \citep{robertBayesianChoice}. The choice of the two constants $k_1$ and $k_2$ is equivalent to the choice of $\kappa = {k_1}/{(k_1 + k_2)}$.

In practice, for some patients we may not have sufficient information to accurately estimate their PSA profile. The resulting high variance of $g(T^*_j)$ could make using a measure of central tendency such as mean or median time of GR unreliable (i.e., overshooting the true $T_j^*$ by a big margin). In such occasions, the approach based on dynamic risk of GR could be more robust. This consideration leads us to a hybrid approach, namely, to select $u$ using dynamic risk of GR based approach when the spread of $g(T_j^*)$ is large, while using $E_g(T^*_j)$ or $\mbox{median}(T^*_j)$ when the spread of $g(T_j^*)$ is small. What constitutes a large spread will be application-specific. In PRIAS, within the first 10 years, the maximum possible delay in detection of GR is three years. Thus we propose that if the difference between the 0.025 quantile of $g(T^*_j)$, and $E_g(T^*_j)$ or $\mbox{median}(T^*_j)$ is more than three years then proposals based on dynamic risk of GR be used instead.

%% file: mainmatter/pers_sched_framework/estimation.tex
% !TEX root =  ../../pers_schedules.tex 

\subsection{Estimation}
Since there is no closed form solution available for $E_g(T^*_j)$, for its estimation we utilize the following relationship between $E_g(T^*_j)$ and $\pi_j(u \mid t, s)$:
\begin{equation}
\label{eq : expected_time_survprob}
E_g(T^*_j) = t + \int_t^\infty \pi_j(u \mid t, s) \rmn{d} u.
\end{equation}
There is no closed form solution available for the integral in (\ref{eq : expected_time_survprob}), and hence we approximate it using Gauss-Kronrod quadrature. We preferred this approach over Monte Carlo methods to estimate $E_g(T^*_j)$ from $g(T^*_j)$, because sampling directly from $g(T^*_j)$ involved an additional step of sampling from the distribution $p(T^*_j \mid T^*_j > t, \boldsymbol{b_j}, \boldsymbol{\theta})$, as compared to the estimation of $\pi_j(u \mid t, s)$ \citep{rizopoulos2011dynamic}. The former approach was thus computationally faster. 

As mentioned earlier, selection of the optimal biopsy time based on $E_g(T_j^*)$ alone will not be practically useful when the $\mbox{var}_g(T^*_j)$ is large, which is given by:
\begin{equation}
\label{eq : var_time_survprob}
\mbox{var}_g(T^*_j) = 2 \int_t^\infty {(u-t) \pi_j(u \mid t, s) \rmn{d} u} - \Big\{\int_t^\infty \pi_j(u \mid t, s) \rmn{d} u\Big\}^2.
\end{equation}
Since a closed form solution is not available for the variance expression, it is estimated similar to the estimation of $E_g(T^*_j)$. The variance depends both on last biopsy time $t$ and PSA history $\mathcal{Y}_j(s)$. The impact of the observed information on variance is demonstrated in Section \ref{subsec : demo_prias_pers_schedule}.

For schedules based on dynamic risk of GR, the value of $\kappa$ dictates the biopsy schedule and thus its choice has important consequences. Often it may be chosen on the basis of the amount of risk that is acceptable to the patient. For example, if the maximum acceptable risk is 5\%, then $\kappa = 0.95$.

In cases where $\kappa$ cannot be chosen on the basis of the input of the patients, we propose to automate the choice of $\kappa$. More specifically, we propose to choose a threshold $\kappa$ for which a binary classification accuracy measure \citep{lopez2014optimalcutpoints}, discriminating between cases and controls, is maximized. In PRIAS, cases are patients who experience GR and the rest are controls. However, a patient can be in control group at some time $t$ and in the cases at some future time point $t + \Delta t$, and thus time dependent binary classification is more relevant. In joint models, a patient $j$ is predicted to be a case if $\pi_j(t + \Delta t \mid t,s) \leq \kappa$ and a control if $\pi_j(t + \Delta t \mid t,s) > \kappa$ \citep*{rizopoulosJMbayes, landmarking2017}. In this work we choose the time window $\Delta t$ to be one year. This because, in AS programs at any point in time, it is of interest to identify patients who may obtain GR in the next one year from those who do not, so that they can be provided immediate attention (in exceptional cases a biopsy within an year of the last one). As for the choice of the binary classification accuracy measure, we require a measure which is in line with the goal to focus on patients whose true time of GR falls in the time window $\Delta t$. To this end, a measure which combines both sensitivity and precision is the $\mbox{F}_1$ score. It is defined as:
\begin{align*}
\mbox{F}_1(t, \Delta t, s) &= 2\frac{\mbox{TPR}(t, \Delta t, s)\ \mbox{PPV}(t, \Delta t, s)}{\mbox{TPR}(t, \Delta t, s) + \mbox{PPV}(t, \Delta t, s)},\\
\mbox{TPR}(t, \Delta t, s) &= \mbox{Pr}\big\{\pi_j(t + \Delta t \mid t,s) \leq \kappa \mid t < T^*_i \leq t + \Delta t\big\},\\
\mbox{PPV}(t, \Delta t, s) &= \mbox{Pr}\big\{t < T^*_i \leq t + \Delta t \mid \pi_j(t + \Delta t \mid t,s) \leq \kappa \big\}.
\end{align*}
where $\mbox{TPR}(\cdot)$ and $\mbox{PPV}(\cdot)$ denote time dependent true positive rate (sensitivity) and positive predictive value (precision), respectively. The estimation for both is similar to the estimation of $\mbox{AUC}(t, \Delta t, s)$ given by \citet{landmarking2017}. Since a high $\mbox{F}_1$ score is desired, the optimal value of $\kappa$ is $\argmax_{\kappa} \mbox{F}_1(t, \Delta t, s)$. In this work we compute the latter using a grid search approach. That is, first $\mbox{F}_1$ is computed using the available dataset over a fine grid of $\kappa$ values between 0 and 1, and then $\kappa$ corresponding to the highest $\mbox{F}_1$ is chosen. Furthermore, in this paper we use $\kappa$ chosen only on the basis of $\mbox{F}_1$ score.

%% file: mainmatter/pers_sched_framework/algorithm.tex
\resizebox{\columnwidth}{!}{
\begin{tikzpicture}
\node (start) [startstop_big] {
Enter Active Surveillance.
\begin{enumerate}
\item Measure PSA and Gleason at induction.
\item Reset $s=t=T^{nv}=0$. 
\item Reset $u = u^{pv} = \infty$.
\end{enumerate}
};

\node (propTime) [process_wide_5cm, below=1cm of start] {
(1) Update $g(T^*_j)$.\\
(2) Set $u = \mbox{new optimal } u$.
};

\node (decision1) [decision, below = 1.0cm of propTime] {$u \leq u^{pv}$};
\node (pro6) [process, right = 2.45cm of decision1] {Set $u = u^{pv}$.};

\node (takePSA) [process_wide_4pt5cm, right=1.5cm of propTime] {
(1) Set $s=s^{nv}$.\\
(2) Measure PSA at $s$. 
};

\node (decision5) [decision, left=1.5cm of decision1] {$u \leq s$};

\node (pro5) [process, below=1.5cm of decision5] {Set $u = s$.};

\node (decision2) [decision, below=0.5cm of decision1] {$u - t \geq 1$};

\node (decision4) [decision, right=2.5cm of decision2] {$u > s^{nv}$};

\node (pro8) [process, right=1cm of pro6] {Set $u^{pv}=u$.};

\node (pro3) [process, below=1.0cm of decision2] {Set $u = t + 1$.};

\node (pro4) [process_wide_4cm, below=1cm of decision4] {Conduct biopsy at $u$.};

\node (decision3) [decision, below=0.5cm of pro3] {$\mbox{Gleason} > 6$};
\node (pro7) [process_wide_4pt5cm, left=2.635cm of decision3] {
(1) Set $t = u$.\\
(2) Reset $u = u^{pv}=\infty$.
};

\node (stop) [startstop, below = 1cm of decision3] {Remove patient from AS};

\draw [arrow] (start) -- (propTime);
\draw [arrow] (takePSA) -- (propTime);
\draw [arrow] (propTime) -- (decision1);
\draw [arrow] (decision1) -- node[anchor=south] {Yes} (decision5);
\draw [arrow] (decision5) -- node[anchor=east] {Yes} (pro5);
\draw [arrow] (pro5) -- (decision2);
\draw [arrow] (decision1) -- node[anchor=south] {No} (pro6);
\draw [arrow] (decision5) -- node[anchor=south] {No} (decision2);
\draw [arrow] (pro6) -- (decision4);
\draw [arrow] (decision4.east) |- ([xshift=2.65cm, yshift=-4.025cm]pro6.north east) -- node[anchor=east] {Yes} (pro8);

\draw [arrow] (pro8) |- (takePSA);

\draw [arrow] (decision2) -- node[anchor=south] {Yes} (decision4);
\draw [arrow] (decision2) -- node[anchor=east] {No} (pro3);
\draw [arrow] (pro3) -- (pro4);
\draw [arrow] (pro4) |- (decision3);
\draw [arrow] (decision3) -- node[anchor=east] {Yes} (stop);
\draw [arrow] (decision4) -- node[anchor=east] {No} (pro4);
\draw [arrow] (decision3) -- node[anchor=south]{No} (pro7);
\draw [arrow] (pro7.north)|- ([xshift=-0.375cm, yshift=-5.25cm]pro5.north west) |- (propTime);
\end{tikzpicture}
}

%% file: mainmatter/choosing_schedule.tex
% !TEX root =  ../pers_schedules.tex

\section{Evaluation of Schedules}
\label{sec : choosing_schedule}
Given a particular schedule $S$ of biopsies, our next goal is to evaluate the schedule and to compare it with other schedules. To this end, we first present the methods to evaluate the biopsy schedules and then discuss the choice of a schedule.

We evaluate a schedule $S$ using two criteria, namely the number of biopsies $N^S_j \geq 1$ a schedule conducts for the $j$-th patient to detect GR, and the offset $O^S_j \geq 0$ by which it overshoots the true GR time $T^*_j$. The offset $O^S_j$ is defined as $O^S_j = T^S_{j{N^S_j}} - T^*_j$, where $T^S_{j{N^S_j}} \geq T^*_j$ is the time at which GR is detected. Our interest lies in the joint distribution $p(N^S_j, O^S_j)$ of the number of biopsies and the offset. Given the medical burden of biopsies, ideally only one biopsy with zero offset should be conducted. Hence, realistically we should select a schedule with a low mean number of biopsies $E(N^S_j)$ as well a low mean offset $E(O^S_j)$. It is also desired that a schedule has low variance of the number of biopsies $\mbox{var}(N^S_j)$, as well as low variance of the offset $\mbox{var}(O^S_j)$, so that the schedule works similarly for most patients. Quantiles of $p(N^S_j)$ may also be of interest. For example, it may be desired that a schedule detects GR with at most two biopsies in 95\% of the patients.

\subsection{Choosing a Schedule}
Given the multiple criteria for evaluation of a schedule, the next step is to use them to select a schedule. Using principles from compound optimal designs \citep{lauter1976optimal} we propose to choose a schedule $S$ which minimizes a loss function of the following form:
\begin{equation}
\label{eq : loss_func_sim_study_generic}
L(S) = \sum_{r=1}^R \eta_r \mathcal{R}_r(N^S_j).
\end{equation}
where $\mathcal{R}_r(\cdot)$ is an evaluation criteria based on either the number of biopsies or the offset (for brevity of notation, only $N^S_j$ is used in the equation above). Some examples of $\mathcal{R}_r(\cdot)$ are mean, median, variance and quantile function. Constants $\eta_1, \ldots, \eta_R$, where $0 \leq \eta_r \leq 1$ and $\sum_{r=1}^R \eta_r = 1$, are weights to differentially weigh-in the contribution of each of the $R$ criteria. An example loss function is:
\begin{equation}
\label{eq : loss_func_sim_study}
L(S) = \eta_1 E(N^S_j) + \eta_2 E(O^S_j). 
\end{equation}
The choice of $\eta_1$ and $\eta_2$ is not easy, because biopsies have associated medical risks and consequently the cost of an extra biopsy cannot be quantified or compared to a unit increase in offset easily. To obviate this problem we utilize the equivalence between compound and constrained optimal designs \citep{cook1994equivalence}. More specifically, it can be shown that for any $\eta_1$ and $\eta_2$ there exists a constant $C>0$ for which minimization of loss function in (\ref{eq : loss_func_sim_study}) is equivalent to minimization of the loss function subject to the constraint that $E(O^S_j) < C$. That is, a suitable schedule is the one which conducts the least number of biopsies to detect GR while simultaneously guaranteeing an offset less than $C$. The choice of $C$ now can be based on the protocol of AS program. In the more generic case in (\ref{eq : loss_func_sim_study_generic}), a schedule can be chosen by minimizing $\mathcal{R}_R(\cdot)$ under the constraint $\mathcal{R}_r(\cdot) < C_r; r=1, \ldots, R-1$.

%% file: mainmatter/pers_schedule_prias.tex
% !TEX root =  ../pers_schedules.tex

\section{Demonstration of Personalized Schedules}
\label{sec : pers_schedule_PRIAS}
To demonstrate how the personalized schedules work, we apply them to the patients enrolled in PRIAS. To this end, we divide the PRIAS dataset into a training dataset with 5264 patients and a demonstration dataset with three patients who never experienced GR. We fit a joint model to the training dataset and then use it to create personalized schedules for patients in demonstration dataset. We fit the joint model using the R package \textbf{JMbayes} \citep{rizopoulosJMbayes}, which uses the Bayesian methodology to estimate the model parameters.

\subsection{Fitting the Joint Model to PRIAS Dataset}
\label{subsec : jm_fit_prias}
The training dataset contains age at the time of induction in PRIAS, PSA levels and the time interval in which GR is detected, for 5264 prostate cancer patients. PSA was measured at every three months for the first two years and every six months thereafter. To detect GR, biopsies were conducted as per the PRIAS schedule (Section \ref{sec : introduction}). For the longitudinal analysis of PSA we use $\log_2 \mbox{PSA}$ measurements instead of the raw data \citep{nieboer2015nonlinear}. The longitudinal sub-model of the joint model we fit is given by:
\begin{equation}
\label{eq : long_model_prias}
\begin{aligned}
\log_2 \mbox{PSA}(t) &= \beta_0 + \beta_1 (\mbox{Age}-70) + \beta_2 (\mbox{Age}-70)^2 + \sum_{k=1}^4 \beta_{k+2} B_k(t,\mathcal{K})\\ 
&+  b_{i0} + b_{i1} B_7(t, 0.1) + b_{i2} B_8(t, 0.1) +
\varepsilon_i(t).
\end{aligned}
\end{equation}
where $B_k(t, \mathcal{K})$ denotes the $k$-th basis function of a B-spline with three internal knots at $\mathcal{K} =\{0.1, 0.5, 4\}$ years, and boundary knots at zero and seven years. The spline for the random effects consists of one internal knot at 0.1 years and boundary knots at zero and seven years. The choice of knots was based on exploratory analysis as well as on model selection criteria AIC and BIC. Age of patients was median centered to avoid numerical instabilities during parameter estimation. For the relative risk sub-model the hazard function we fit is given by:
\begin{equation}
\label{eq : hazard_prias}
h_i(t) = h_0(t) \exp\big\{\gamma_1 (\mbox{Age}-70)  + \gamma_2 (\mbox{Age}-70)^2 + \alpha_1 m_i(t) + \alpha_2 m'_i(t)\big\}.
\end{equation}
where $\alpha_1$ and $\alpha_2$ are measures of strength of the association between hazard of GR and $\log_2 \mbox{PSA}$ value $m_i(t)$ and $\log_2 \mbox{PSA}$ velocity $m'_i(t)$, respectively. Since the PRIAS schedule depends only on the observed PSA values (via PSA-DT), the interval censoring observed in PRIAS is independent and non informative of the underlying health of the patient.

From the joint model fitted to the PRIAS dataset we found that only $\log_2 \mbox{PSA}$ velocity was strongly associated with hazard of GR. For any patient, a unit increase in $\log_2 \mbox{PSA}$ velocity led to an 11 fold increase in the hazard of GR. The parameter estimates for the fitted joint model are presented in detail in Web Appendix C of the supplementary material. 

\subsection{Personalized Schedules for the First Demonstration Patient}
\label{subsec : demo_prias_pers_schedule}
\begin{figure}
\centerline{
\includegraphics[width=\columnwidth]{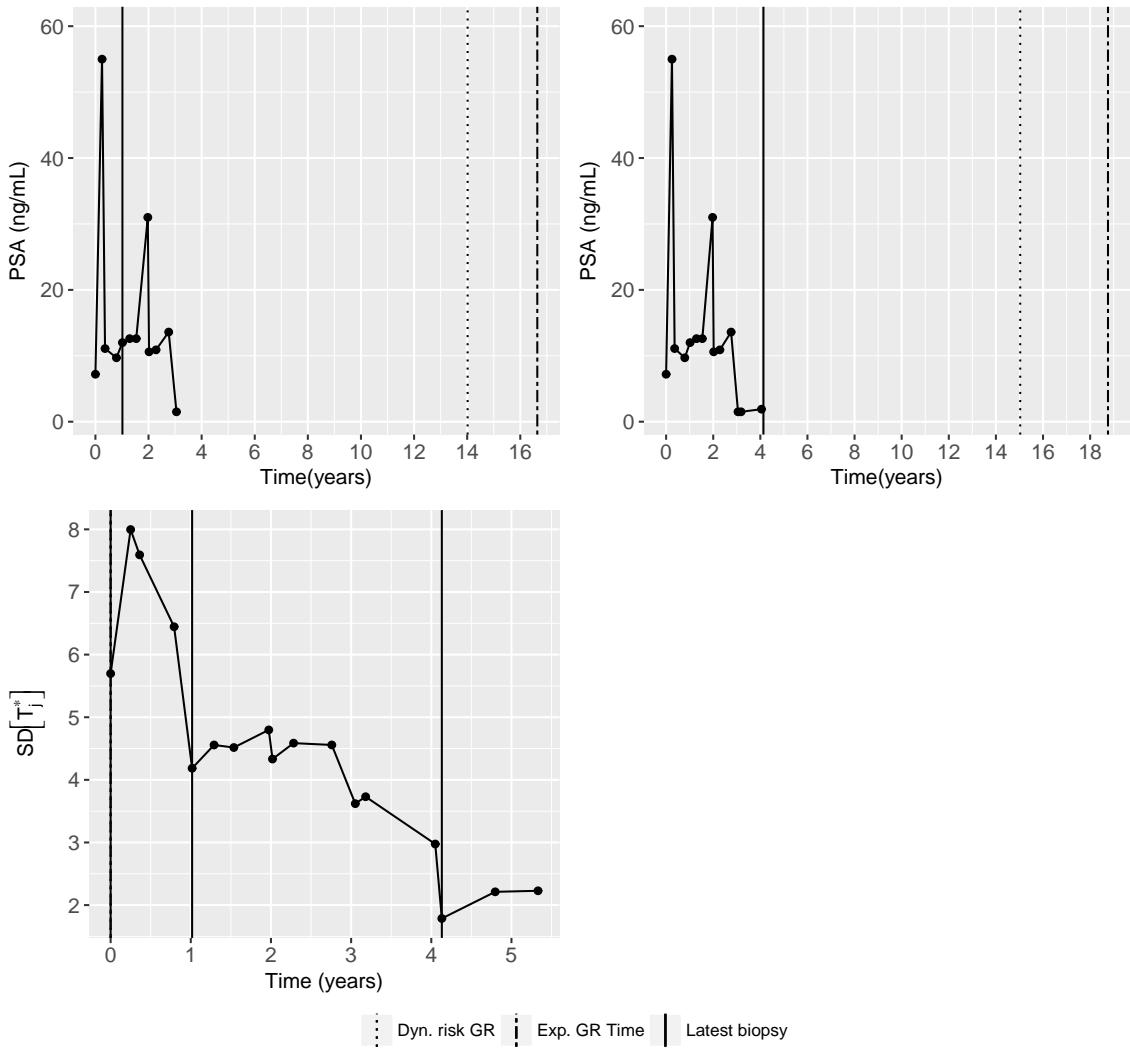}
}
\caption{Top panel: Evolution of PSA, history of repeat biopsies and corresponding personalized schedules for the first demonstration patient. Bottom Panel: History of repeat biopsies and $\mbox{SD}_g(T^*_j) = \sqrt{\mbox{var}_g(T^*_j)}$ over time for the first demonstration patient.}
\label{web_fig : prias_demo_pid_911}
\end{figure}

Using the demonstration dataset, we next present the functioning of personalized schedules based on expected time of GR and dynamic risk of GR. The evolution of PSA, time of last biopsy and proposed biopsy times for the first demonstration patient are shown in the top panel of Figure \ref{web_fig : prias_demo_pid_911}. We can see the combined effect of decreasing PSA levels and a negative repeat biopsy on personalized schedules, between year three and year 4.5 for this patient. In accordance with the two negative repeat biopsies and consistently decreasing PSA, the proposed time of biopsy based on dynamic risk of GR increases from 14 years to 15 years in this period. Whereas, the proposed time of biopsy based on expected time of GR increases from 16.6 years to 18.8 years. We can also see in the bottom panel of Figure \ref{web_fig : prias_demo_pid_911} that after each negative repeat biopsy, $\mbox{SD}[T^*_j] = \sqrt{\mbox{var}_g(T^*_j)}$ decreases sharply. Thus, if the expected time of GR based approach is used, then the offset $O^S_j$ will be smaller on average for biopsies scheduled after the second repeat biopsy than those scheduled after the first repeat biopsy.

The demonstration of personalized schedules for the two other patients from the demonstration data set is presented in Web Appendix D of the supplementary material.

%% file: mainmatter/sim_study.tex
% !TEX root =  ../pers_schedules.tex 
\section{Simulation Study}
\label{sec: simulation_study}
The application of personalized schedules for patients from PRIAS demonstrated that these schedules adapt according to the historical data of each patient. However we could not perform a full scale comparison between personalized and PRIAS schedule, because the true time of GR was not known for the PRIAS patients. To this end, we conducted a simulation study comparing personalized schedules with PRIAS and annual schedule, whose details are presented next.

\subsection{Simulation Setup}
\label{subsec : simulation_setup}
First we assume a population of patients enrolled in AS, with the same entrance criteria as that of PRIAS. The PSA and hazard of GR for patients from this population follow a joint model of the form postulated in Section \ref{subsec : jm_fit_prias}, with parameters equal to the posterior mean of parameters estimated from the joint model fitted to PRIAS dataset (Web Appendix C of the supplementary material). We further assume that there are three equal sized subgroups $G_1$, $G_2$ and $G_3$ of patients in the population, differing in the baseline hazard of GR. This was done because we wanted to test the performance of different schedules for a population with a mixture of patients, namely those with faster progressing PCa, as well as those with slowly progressing PCa. For the three subgroups we use a Weibull distributed baseline hazard with the following shape and scale parameters $(k, \lambda$): $(1.5, 4)$, $(3, 5)$ and $(4.5, 6)$ for $G_1, G_2$ and $G_3$, respectively. The effect of these parameters is that the mean GR time is lowest in $G_1$ (faster progressing PCa) and highest in $G_3$ (slowly progressing PCa).

From this population we have sampled 500 datasets with 1000 patients each. Patients are randomly assigned to a subgroup. Further, each dataset is split into a training (750 patients) and a test (250 patients) part. The $k$-th simulated training dataset $\mathcal{D}^k$ is given by $\mathcal{D}^k = \{l_{ki}, r_{ki}, \boldsymbol{y}_{ki}; i = 1, \ldots, 750\}$, where $\boldsymbol{y}_{ki}$ denote the PSA measurements for the $i$-th patient in $\mathcal{D}^k$. The frequency of PSA measurements is same as that in PRIAS. Other than simulating a true GR time $T^*_{ki}$, we also generate a random and non-informative censoring time $C_{ki}$. When $T^*_{ki} < C_{ki}$, then $l_{ki} = r_{ki} = T^*_{ki}$, otherwise $l_{ki} = C_{ki}$ and $r_{ki} = \infty$. For the test patients, censoring time is not generated.

We next fit a joint model of the specification given in (\ref{eq : long_model_prias}) and (\ref{eq : hazard_prias}) to each of the $\mathcal{D}^k, k=1, \ldots, 500$, and obtain a MCMC sample from the posterior distribution $p(\boldsymbol{\theta} \mid \mathcal{D}^k)$. We then obtain $g(T^*_{kl})$ for each of the $l$-th test patient of the $k$-th data set and conduct hypothetical biopsies for him. For every patient we conduct biopsies using the following six types of schedules (abbreviated names in parenthesis): personalized schedules based on expected time of GR (Exp. GR time) and median time of GR (Med. GR time), personalized schedules based on dynamic risk of GR (Dyn. risk GR), a hybrid approach between median time of GR and dynamic risk of GR (Hybrid), PRIAS schedule and annual schedule. The biopsies are conducted iteratively in accordance with the algorithm in Figure \ref{fig : sched_algorithm}. 

To compare the aforementioned schedules we require estimates of the various criteria based on offset and number of biopsies conducted to detect GR (Section \ref{sec : choosing_schedule}). To this end, we compute pooled estimates of each of the $E(N^S_j)$, $\mbox{var}(N^S_j)$, $E(O^S_j)$ and $\mbox{var}(O^S_j)$, as below:
\begin{align*}
\widehat{E(O^S_j)} &= \frac{\sum_{k=1}^{500} n_k \widehat{E(O^S_k)}}{\sum_{k=1}^{500} n_k}, \\
\widehat{\mbox{var}(O^S_j)} &= \frac{\sum_{k=1}^{500} (n_k - 1) \widehat{\mbox{var}(O^S_k)}}{\sum_{k=1}^{500} (n_k-1)}, 
\end{align*}
where $n_k$ denotes the number of test patients, $\widehat{E(O^S_k)} = {\sum_{l=1}^{n_k}O^S_{kl}}/{n_k}$ is the estimated mean and $\widehat{\mbox{var}(O^S_k)} = {\sum_{l=1}^{n_k}\big\{O^S_{kl} - \widehat{E(O^S_k)}\big\}^2}/(n_k-1)$ is the estimated variance of the offset for the $k$-th simulation. The estimates for number of biopsies are obtained similarly.

\subsection{Results}
\begin{figure}
\centerline{\includegraphics[width=\columnwidth]{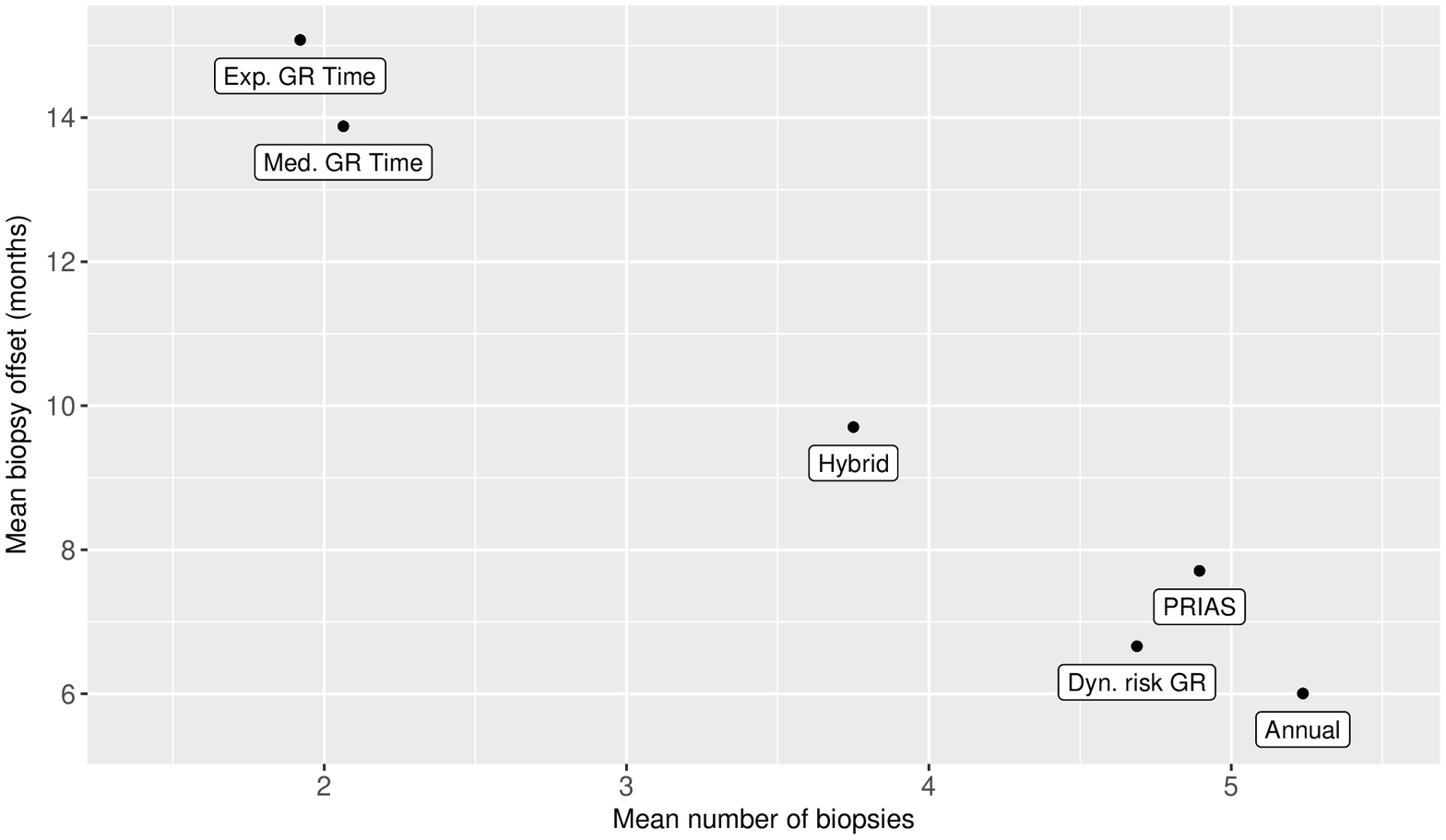}}
\caption{Estimated mean number of biopsies and mean offset (months) for the 6 schedules, using all simulated patients.}
\label{fig : meanNbVsOffset}
\end{figure}

The pooled estimates of the aforementioned criteria are summarized in Table \ref{table : sim_study_pooled_estimates}. In addition, mean offset is plotted against mean number of biopsies conducted to detect GR in Figure \ref{fig : meanNbVsOffset}. From the figure it is evident that across the schedules there is an inverse relationship between $E(N^S_j)$ and $E(O^S_j)$. For example, the annual schedule conducts on average 5.2 biopsies to detect GR, which is the highest among all schedules, however it has the least average offset of 6 months as well. On the other hand the schedule based on expected time of GR conducts only 1.9 biopsies on average to detect GR, the least among all schedules, but it also has the highest average offset of 15 months. The schedule based on median time of GR performs similar to that based on expected time of GR. Since the annual schedule attempts to contain the offset within an year it has the least $\mbox{SD}(O^S_j) = \sqrt{\mbox{var}(O^S_j)}$. However to achieve so, it conducts a wide range of number of biopsies from patient to patient, i.e., highest $\mbox{SD}(N^S_j) = \sqrt{\mbox{var}(N^S_j)}$. Schedules based on expected and median time of GR perform the opposite of annual schedule in terms of $\mbox{SD}(N^S_j)$ and $\mbox{SD}(O^S_j)$.

%124781 = 41484 + 41423 + 41874
\begin{table}
\caption{Estimated mean and standard deviation of the number of biopsies and offset (months).}
\label{table : sim_study_pooled_estimates}
\begin{tabular}{lrrrr}
\Hline
\multicolumn{5}{c}{a) All subgroups}\\
\hline
Schedule          & $E(N^S_j)$ & $E(O^S_j)$ & ${\mbox{SD}(N^S_j)}$ & ${\mbox{SD}(O^S_j)}$ \\
\hline
Annual         & 5.24            & 6.01                & 2.53          & 3.46              \\
PRIAS          & 4.90            & 7.71                & 2.36          & 6.31\\
Dyn. risk GR       & 4.69            & 6.66                & 2.19           & 4.38              \\
Hybrid       & 3.75            & 9.70                & 1.71          & 7.25              \\
Med. GR time & 2.06            & 13.88               & 1.41          & 11.80              \\
Exp. GR time & 1.92            & 15.08               & 1.19          & 12.11             \\
\hline
\multicolumn{5}{c}{b) Subgroup $G_1$}\\
\hline
Schedule        & $E(N^S_j)$ & $E(O^S_j)$ & ${\mbox{SD}(N^S_j)}$ & ${\mbox{SD}(O^S_j)}$ \\
\hline
Annual         & 4.32            & 6.02                & 3.13          & 3.44              \\
PRIAS          & 4.07            & 7.44                & 2.88          & 6.11    \\
Dyn. risk GR       & 3.85            & 6.75                & 2.69          & 4.44              \\
Hybrid       & 3.25            & 10.25               & 2.16          & 8.07              \\
Med. GR time & 1.84            & 20.66               & 1.76          & 14.62             \\
Exp. GR time & 1.72            & 21.65               & 1.47          & 14.75             \\
\hline      
\multicolumn{5}{c}{c) Subgroup $G_2$}\\
\hline
Schedule        & $E(N^S_j)$ & $E(O^S_j)$ & ${\mbox{SD}(N^S_j)}$ & ${\mbox{SD}(O^S_j)}$ \\
\hline
Annual         & 5.18            & 5.98                & 2.13          & 3.47              \\
PRIAS          & 4.85            & 7.70                & 2.00          & 6.29        \\
Dyn. risk GR       & 4.63            & 6.66                & 1.82          & 4.37              \\
Hybrid       & 3.68            & 10.32                & 1.37          & 7.45              \\
Med. GR time & 1.89             & 12.33               & 1.16          & 9.44              \\
Exp. GR time & 1.77            & 13.54               & 0.98          & 9.83              \\
\hline      
\multicolumn{5}{c}{d) Subgroup $G_3$}\\
\hline
Schedule        & $E(N^S_j)$ & $E(O^S_j)$ & ${\mbox{SD}(N^S_j)}$ & ${\mbox{SD}(O^S_j)}$ \\
\hline
Annual         & 6.20             & 6.02                & 1.76          & 3.46              \\
PRIAS          & 5.76             & 7.98                & 1.71         & 6.51        \\
Dyn. risk GR       & 5.58            & 6.58                & 1.56          & 4.33              \\
Hybrid       & 4.32            & 8.55                & 1.26          & 5.91              \\
Med. GR time & 2.45            & 8.70                & 1.15          & 6.32              \\
Exp. GR time & 2.27            & 10.09               & 0.99          & 7.47              \\
\hline     
\end{tabular}
\end{table}

The PRIAS schedule conducts only 0.3 biopsies less than the annual schedule, but with a higher variance of offset, it does not guarantee early detection for everyone. If we compare the PRIAS schedule with dynamic risk of GR based schedule, we can see that the latter performs slightly better than PRIAS schedule in all four criteria. The hybrid approach combines the benefits of methods with low $E(N^S_j)$ and $\mbox{SD}(N^S_j)$, and methods with low $E(O^S_j)$ and $\mbox{SD}(O^S_j)$. It conducts 1.5 biopsies less than the annual schedule on average and with a $E(O^S_j)$ of 9.7 months it detects GR within an year since its occurrence. Moreover, it has both $\mbox{SD}(N^S_j)$ and $\mbox{SD}(O^S_j)$ comparable to PRIAS.

The performance of each schedule differs for the three subgroups $G_1, G_2$ and $G_3$. The annual schedule remains the most consistent across subgroups in terms of the offset, but it conducts 2 extra biopsies for subgroup $G_3$ (slowly progressing PCa) than $G_1$ (faster progressing PCa). The performance of schedule based on expected time of GR is the most consistent in terms of number of biopsies but it detects GR an year later on average in subgroup $G_1$ than $G_3$. For the dynamic risk of GR based schedule and the hybrid schedule the dynamics are similar to that of the annual schedule. Unlike the latter two schedules, the PRIAS schedule not only conducts more biopsies in $G_3$ than $G_1$ but also detects GR later in $G_3$ than $G_1$.

\begin{figure}[!htb]
\centerline{\includegraphics[width=\columnwidth]{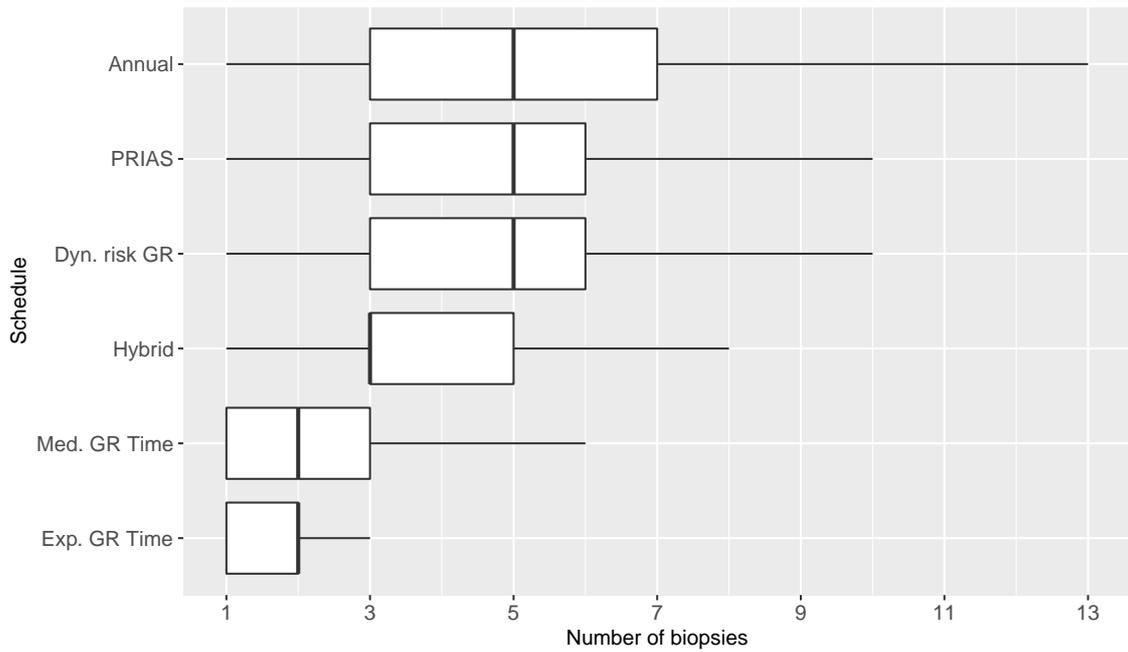}}
\caption{Boxplot showing variation in number of biopsies conducted by different methods, using all simulated patients.}
\label{fig : nbBoxPlot_all}
\end{figure}

\begin{figure}[!htb]
\centerline{\includegraphics[width=\columnwidth]{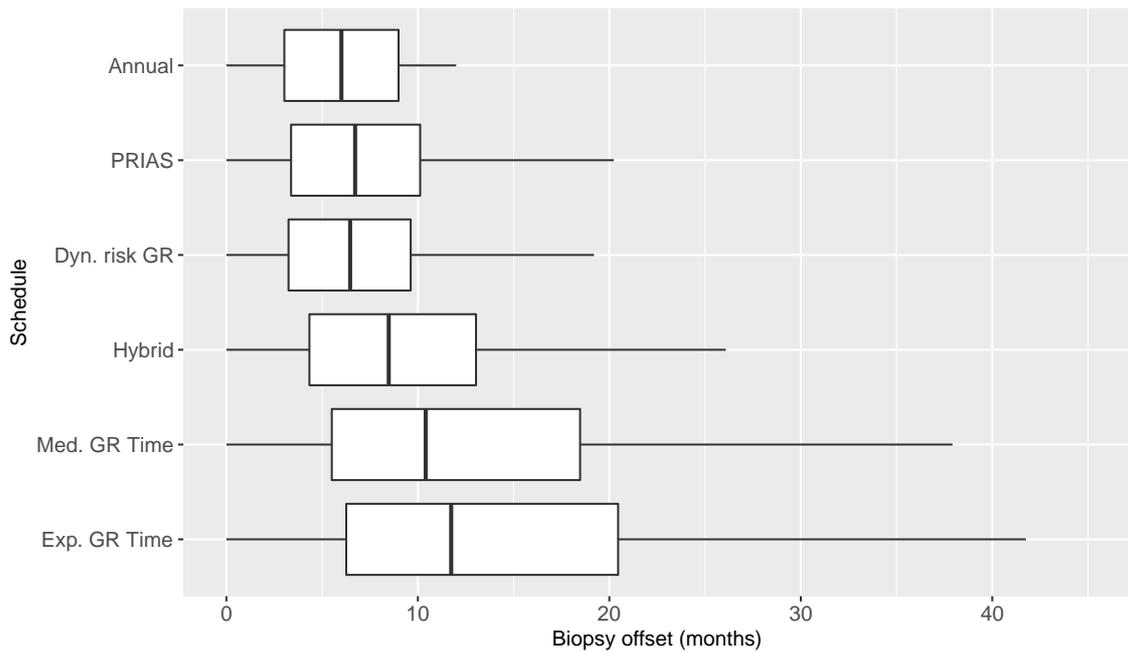}}
\caption{Boxplot showing variation in biopsy offset (months) for different methods, using all simulated patients.}
\label{fig : offsetBoxPlot_all}
\end{figure}

The choice of a suitable schedule using (\ref{eq : loss_func_sim_study_generic}) depends on the chosen criteria for evaluation of schedules. For example, the schedule based on dynamic risk of GR is suitable if on average the least number of biopsies are to be conducted to detect GR, while simultaneously making sure that at least 90\% of the patients have an average offset less than one year (Figure \ref{fig : nbBoxPlot_all} and \ref{fig : offsetBoxPlot_all}). The schedule based on expected time of GR is suitable if on average the least number of biopsies are to be conducted to detect GR, while simultaneously making sure that at least 90\% of the patients have an average offset less than three years. If a stricter cutoff is required on offset the hybrid approach may be suitable, since it conducts only 3.8 biopsies on average while guaranteeing an offset of two years for 95\% of the patients and three years for 99.9\% of the patients. Besides if further cutoffs are required on variance of number of biopsies or offset they are not too high either for the hybrid approach.

%% file: mainmatter/discussion.tex
% !TEX root =  ../pers_schedules.tex 

\section{Discussion}
\label{sec: discussion}
In this paper we presented personalized schedules for surveillance of PCa patients. The problem at hand was the following: Low risk PCa patients enrolled in AS have to undergo repeat biopsies on a frequent basis for examination of PCa progression, and since biopsies have an associated risk of complications, not all patients comply with the schedule of biopsies. This may reduce the effectiveness of AS programs because detection of PCa progression is delayed. To approach these problems we proposed personalized schedules based on joint models for time to event and longitudinal data. At any given point in time, the proposed personalized schedules utilize a patient's information from historical PSA measurements and repeat biopsies conducted up to that time. We proposed two different classes of personalized schedules, namely schedules based on expected and median time of GR of a patient, and schedules based on dynamic risk of GR. In addition we also proposed a combination (hybrid approach) of these two approaches, which is useful in scenarios where variance of time of GR for a patient is high. We then proposed criteria for evaluation of various schedules and a method to select a suitable schedule.

We demonstrated using the PRIAS dataset that the personalized schedules adjust the time of biopsy on the basis of results from historical PSA measurements and repeat biopsies, even when the two are not in concordance with each other (Web Appendix D). Secondly, we conducted a simulation study to compare various schedules. We observed that the schedules based on expected and median time of GR conduct only two biopsies on average to detect GR, which is promising compared to PRIAS (4.9 biopsies) and annual schedule (5.2 biopsies). We also observed that the performance of the schedules depends on the true GR time of the patient. For example, in simulated patients who have a slowly progressing PCa (subgroup $G_3$), personalized schedule based on expected time of GR detects GR one year earlier on average compared to patients who have a faster progressing PCa (subgroup $G_1$), while conducting approximately the same number of biopsies for both subgroups. For subgroup $G_1$, the annual or PRIAS schedule may be preferred because they detect GR at 6 and 7.4 months since its occurrence, respectively. However for slowly progressing PCa patients up to 6 biopsies were needed to detect GR at 6 and 8 months, respectively. In such scenarios, that is, where it is not known in advance if the patient will have a faster or slower progression of PCa, the hybrid approach provides an interesting alternative. This because, it conducts one biopsy less than the annual schedule in faster progressing PCa patients while detecting GR at 10.3 months since its occurrence on average. Whereas, for slowly progressing PCa patients it conducts two biopsies less than the annual schedule while detecting GR at 8.6 months since its occurrence on average.

While each of the personalized schedules have their own advantages and disadvantages, they also offer multiple choices to the AS programs to choose one as per their requirements, instead of choosing a common fixed schedule for all patients. In this regard, we proposed to choose schedules which conduct as less biopsies as possible while restricting the offset to be below a AS program specific threshold, or vice versa. There is also potential to develop more personalized schedules. For example, using loss functions which asymmetrically penalize overshooting/undershooting the target GR time can be interesting. Furthermore, for dynamic risk of GR based schedules, it is also possible to choose $\kappa$ using other binary classification accuracy measures or as per patient's wish (Web Appendix E). Although in this work we assumed that the time of GR was interval censored, in reality the Gleason scores are susceptible to inter-observer variation \citep{Gleason_interobs_var}. Models and schedules which account for error in measurement of time of GR will be interesting to investigate further. Lastly, there is potential for including diagnostic information from magnetic resonance imaging (MRI) or DRE. Unlike PSA levels, such information may not always be continuous in nature, in which case our proposed methodology can be extended by utilizing the framework of generalized linear mixed models.

%% file: mainmatter/acknowledgement.tex
% !TEX root =  ../pers_schedules.tex

\section*{Acknowledgements}
The first and last authors would like to acknowledge support by the Netherlands Organization for Scientific Research's VIDI grant nr. 016.146.301, and Erasmus MC funding. The authors also thank the Erasmus MC Cancer Computational Biology Center for giving access to their IT-infrastructure and software that was used for the computations and data analysis in this study. Lastly, we thank Frank-Jan H. Drost from the Department of Urology, Erasmus University Medical Center, for helping us in cleaning the PRIAS data set. \vspace*{-8pt}